# Thermoelectric anisotropy in Ba(Fe$_{1-x}$Co$_x$)$_2$As$_2$ iron-based superconductor

Marcin Matusiak[1,*], Krzysztof Rogacki[1], Thomas Wolf[2]

*1. Institute of Low Temperature and Structure Research, Polish Academy of Sciences, ul. Okolna 2, 50-422 Wroclaw, Poland*

*2. Institute of Solid State Physics (IFP), Karlsruhe Institute of Technology, D-76021, Karlsruhe, Germany*

We report the in-plane anisotropy of the Seebeck and Nernst coefficients as well as of the electrical resistivity determined for the series of the strain-detwinned single crystals of Ba(Fe$_{1-x}$Co$_x$)$_2$As$_2$. Two underdoped samples ($x$ = 0.024, 0.045) exhibiting the transition from the tetragonal paramagnetic phase to the orthorhombic spin density wave (SDW) phase (at $T_{tr}$ = 100 and 60 K, respectively) show an onset of the Nernst anisotropy at temperatures above 200 K, which is significantly higher than $T_{tr}$. In the optimally doped sample ($x$ = 0.06) the transport properties also appear to be in-plane anisotropic below $T \approx$ 120 K, despite the fact that this particular composition does not show any evidence of long-range magnetic order. However, the anisotropy observed in the optimally doped crystal is rather small and for the Seebeck and Nernst coefficients the difference between values measured along and across the uniaxial strain has opposite sign to those observed for underdoped crystals with $x$ = 0.024 and 0.045. For these two samples, insensitivity of the Nernst anisotropy to the SDW transition suggests that the nematicity might be of other than magnetic origin.

* m.matusiak@int.pan.wroc.pl

**Introduction**

The explanation of an unconventional behaviour of high-temperature superconductors has turned out to be a challenging task for the scientific community [1]. One of early proposed scenarios leading to the development of both high superconducting critical temperatures as well as strange metallic phase above the superconducting dome was related to the presence of a quantum critical point (QCP) [2]. It is worth noting that the occurrence of such QCP was reported for both copper-based [3] and iron-based [4] superconductors. In the latter group of materials the quantum transition is suggested to be related to somewhat enigmatic nematicity [4,5] which in condensed matter is defined as the breaking of rotational symmetry of the electronic system due to correlations rather than the anisotropy of the underlying crystal lattice [6]. Recent studies indicate that the relation between nematicity and superconductivity might be of more general nature [7].

In this Rapid Communication we investigate the in-plane anisotropy of the Seebeck and Nernst coefficients, since the thermoelectric phenomena are known to be sensitive to nematic characteristics of an electronic system [8 - 10]. The thermoelectric response measured for the $Ba(Fe_{1-x}Co_x)_2As_2$ series appears to be significantly different between configurations where the thermal gradient is applied along or across the uniaxial strain. This difference is observed in both tetragonal and orthorhombic phase. The optimally doped sample exhibits the smallest anisotropy and its sign is opposite to that found for the underdoped samples. Our results indicate that the Dirac cone, presumably present in the electronic structure of the $BaFe_2As_2$ parent compound [11 – 13], is effectively eradicated by small cobalt doping. In addition, the determined Nernst anisotropy provides hints about an origin of the nematic fluctuations.

**Experiment**

The $Ba(Fe_{1-x}Co_x)_2As_2$ crystals were grown from self-flux in glassy carbon or alumina crucibles. Particularly low cooling rates of 0.20 to 0.61°C/h were applied to minimize the amount of flux inclusions and crystal defects [14]. The composition of the crystals were determined by energy dispersive x-ray spectroscopy. The series studied in this work consists of four samples with the cobalt content: 0.0 at% (Co0), 2.4 at% (Co2), 4.5 at% (Co4), and 6.0 at% (Co6). For the experiment, square shaped samples were cut out from as grown plate-like single crystals with edges rotated by 45 degree in relation to the tetragonal axes. The sides of the square were about $2 - 2.5$ mm and its thickness $0.1 - 0.3$ mm.

Firstly, the Hall coefficient was measured in unstrained crystals in a magnetic field of $B = 12.5$ T. Then a sample was mounted between two clamps made of phosphor bronze and subjected to a uniaxial pressure applied along its sides by a beryllium copper spring controlled with a stepper motor. For the resistivity ($\rho$) measurements, the electrical contacts were placed at the corners of a sample and the orientations of the voltage and current leads were switched repetitively during the experiment. This allowed the electrical resistivities $\rho_a$ and $\rho_b$ to be determined using the Montgomery method [15]. The uniaxial pressure was increased step-by-step and measurements of the resistivity were repeated until a saturation of the anisotropy, indicating maximal detwinning, was achieved. The maximal pressure determined in this way was used during subsequent thermoelectric experiments.

The Seebeck ($S$) and Nernst ($\nu$) coefficients along and across the strain direction were measured in two separate runs with the magnetic field (parallel to $c$-axis) varied from -12.5 T to +12.5 T. The temperature difference along a sample was determined using two Cernox thermometers and a pre-calibrated in magnetic field constantan – chromel thermocouple attached to the sample through a few millimetres long and 100 μm thick silver wires. Signal leads were made up of long pairs of 25 μm phosphor bronze wires. More details about the experimental setup are given in the Supplemental Material [16].

**Results and discussion**

An expected consequence of the transition from the paramagnetic to the spin density wave (SDW) phase is a change of electronic transport properties due to Fermi surface reconstruction [16]. One of the quantities likely affected by this reconstruction is the Hall coefficient ($R_H$) which in the Ba(Fe$_{1-x}$Co$_x$)$_2$As$_2$ series undergoes a step-like anomaly presented in Fig. 1. The onset of this anomaly was used to determine temperature of the transition at $T_{tr} \approx$ 140, 100, and 60 K in Co0 [17], Co2 and Co4, respectively. The $R_H(T)$ dependence in Co6 exhibits a slight downturn below $T \approx 40$ K that might be a sign of the oncoming SDW transition but instead the system goes directly into the superconducting state. The resistive superconducting transition in Co6 is shown in Figure 2 which presents the temperature dependences of the electrical resistivity for all samples from the Ba(Fe$_{1-x}$Co$_x$)$_2$As$_2$ series detwinned by uniaxial pressure. Two of the samples are superconducting (Co4 with $T_c = 20.3$ K and Co6 with $T_c = 24.5$ K), three of them show the anomaly connected to the structural/magnetic transition (Co0, Co2 and Co4). As previously reported, the resistivities measured along $a$ (long) and $b$ (short) orthorhombic axes ($\rho_a$ and $\rho_b$, respectively) begin do diverge at temperature significantly higher than the temperature of the actual structural transition [18]. Remarkably, in the cobalt doped samples $\rho_a$, which in the orthorhombic phase is smaller than $\rho_b$, seems to be almost unaffected by the transition. The anomaly at $T_{tr}$ is barely noticeable in $\rho_a(T)$ for Co2 and Co4, while for Co6 $\rho_a(T)$ stays linear down to the superconducting transition, in contrast to $\rho_b(T)$ which exhibits an upturn below $T \approx 50$ K. A possible explanation of this behaviour involves the strongly anisotropic scattering of charge carriers [19] influencing only the $b$-axis electronic transport.

This kind of a $b$-axis-only response is not limited to the electrical resistivity but was also observed in the thermoelectric power ($S$), which in EuFe$_2$(As$_{1-x}$P$_x$)$_2$ exhibits a sizeable anomaly at $T_{tr}$ only for measurements along $b$-axis [10]. Temperature dependences of the thermoelectric power for Ba(Fe$_{1-x}$Co$_x$)$_2$As$_2$ presented in Fig. 3 follow the same rule, i.e. anomalies at $T_{tr}$ in Co2 and Co4 are more pronounced in $S_b(T)$ than in $S_a(T)$, however they are still much smaller than those observed in EuFe$_2$(As$_{1-x}$P$_x$)$_2$ [10]. Another similarity between our data and those reported by Jiang et al. [10] is that $S_a$ and $S_b$, which are normalized in the higher temperature limit by multiplying by a small (0.9 − 1.1) correction factor, begin to diverge at a temperature much higher than $T_{tr}$. Figure 4 presents the $\Delta S = S_b - S_a$ difference, which is divided by $T$ to account for the fact that the thermoelectric power, being a measure of entropy per charge carrier [20], decreases with decreasing temperature and $S$ has to drop to zero for $T \rightarrow 0$ K. $\Delta S$ in Ba(Fe$_{1-x}$Co$_x$)$_2$As$_2$ shows an anomaly at $T_{tr}$ and in the entire temperature range is negative for Co2 and Co4, while positive in Co6. In Ref. [10] the inversion of $\Delta S$ at the structural transition ($T_s$) was interpreted as a result of change in the mechanism causing nematicity – from scattering for $T > T_s$ to orbital polarization for $T < T_s$, but we do not observe an equivalent phenomenon at $T_{tr}$ in our series. Hence, in our opinion the change of the $\Delta S$ sign caused by cobalt doping is a consequence of the multiband structure of the 122 iron pnictides [21]. If more than one band takes part in the electronic transport then the effective coefficients are sums of contributions from different bands. For the thermoelectric power this has a form: $S = \frac{\sum_i S_i \sigma_i}{\sum_i \sigma_i}$, where $S_i$ and $\sigma_i$ are, respectively, the thermoelectric power and electrical conductivity of a given $i$-band. The cobalt doping supplies the BaFe$_2$As$_2$ system with additional electrons and can alter both $S_i$ and $\sigma_i$. It is worth noting that even if the character of the electrical resistivity anisotropy does not change ($\rho$ along the strain remains higher than $\rho$ across the strain), the anisotropy of the thermoelectric power still can reverse. This can happen in an electronic system consisting of isotropic and anisotropic bands, when the ratio between respective $S$ coefficients changes disproportionately due to a shift of the Fermi level. An analogous scenario was proposed to explain the anisotropy of the Nernst coefficient ($\nu$) in the SDW phase of BaFe$_2$As$_2$ and CaFe$_2$As$_2$ [22].

As shown in Fig. 5 the magnetic/structural transition in the parent compound Co0 causes the Nernst coefficient to rise dramatically in a way similar to one reported for CaFe$_2$As$_2$ [23] and EuAs$_2$Fe$_2$ [24]. In contrast, a small cobalt substitution is sufficient to inverse this anomaly, namely, $\nu$ in Co2 (2.4 at% cobalt content) decreases below $T_{tr}$ and changes sign to negative below $T \approx 60$ K. Such kind of response of the Nernst effect to cobalt doping was observed in Eu(Fe$_{1-x}$Co$_x$)$_2$As$_2$ [24], where the sudden change of the character of the anomaly was attributed to the suppression of the Dirac band contribution, perhaps due to a shift of the Fermi level or an enhanced scattering of Dirac fermions. Interestingly, unlike temperature dependences of the resistivity and thermoelectric power, $\nu(T)$ measured along $a$-axis and $b$-axis ($\nu_a$ and $\nu_b$, respectively) below $T_{tr}$ look quite similar for each of the samples. This is likely related to the fact that the Nernst coefficient measured along $a$-direction is subject to both $a$- and $b$-direction scattering processes:

$$\nu_x B = -\frac{\alpha_{yx}}{\sigma_{yy}} - S_x R_H \sigma_{xx} , \qquad (1)$$

($\alpha_{yx}$ is the off-diagonal element of the Peltier tensor, $\sigma_{xx}$ and $\sigma_{yy}$ are diagonal elements of the electrical conductivity tensor).

The anisotropy of the Nernst coefficient defined as a difference between $\nu_b$ and $\nu_a$ ($\Delta\nu = \nu_b\text{-}\nu_a$) divided by $T$ is plotted versus temperature in Fig. 6. The first appearance of the Nernst anisotropy is noticeable at temperatures $T_f \approx$ 215, 225, 205, and 120 K for Co0, Co2, Co4, and Co6, respectively, which are significantly higher than $T_{tr}$. These values are in good agreement with the onset temperatures of nematic fluctuations estimated from Raman spectroscopy by Kretzschmar et al. [25]. Another similarity between our results and those reported in Ref. [25] is that a sizeable nematic behaviour is observed for Ba(Fe$_{1-x}$Co$_x$)$_2$As$_2$ in a rather narrow range of Co-doping. Namely, Kretzschmar et al. detected the nematic response up to $x = 0.061$ and they called the effect "unambiguous" only up to $x = 0.051$. Correspondingly, we see that our sample Co6 ($x = 0.06$) is one that behaves differently from Co0, Co2 and Co4, i.e. the anisotropy of the Nernst coefficient in Co6 is small, emerges at lower temperature, and most importantly, the sign of $\Delta\nu$ for this sample, analogously to $\Delta S$, is inversed. Here again, we attribute this inversion to an interplay between different conductivity bands. $\Delta\nu/T$ decays exponentially with temperature ($\Delta\nu(T)/T \sim e^{-cT}$) with the parameter $c = 3.3$ x $10^{-2}$ for Co0 and $c = 1.8$ x $10^{-2}$ for both Co2 and Co4. Such an exponential temperature dependence of the Nernst anisotropy might be surprising, since the nematic susceptibility was reported to obey a Curie-Weiss power law in Ba(Fe$_{1-x}$Co$_x$)$_2$As$_2$ as well as in other optimally doped iron-based superconductors [4].

Possibly the most unexpected outcome of the present studies is that we do not observe any anomaly in $\Delta\nu/T$ at the magnetic/structural transition neither in Co2 nor in Co4. This means that presumed scattering processes that cause rise of the anisotropy at $T_{tr}$ in $\rho$ and $S$, apparently by affecting transport along $b$-axis, cancel out for the Nernst anisotropy since it consists of contributions from both $a$- and $b$-direction transport. Perhaps it is related to the form of $\Delta\nu$, which can be expressed as a subtraction of a two Sondheimer cancellations [26], one for $a$-direction, another one for $b$-direction: $\Delta\nu = \left[\left(\frac{\alpha_{xy}}{\sigma_{xx}} - S_x R_H \sigma_{xx}\right) - \left(\frac{\alpha_{yx}}{\sigma_{yy}} - S_y R_H \sigma_{yy}\right)\right]/B$. It is noteworthy that the slope $c$ of the $\Delta\nu(T)/T \sim e^{-cT}$ dependences in Co2 and Co4 stays the same in both the paramagnetic and in the SDW phase, which suggests that the nematic fluctuations responsible for the high temperature anisotropy of the Nernst coefficient are not affected by formation of the magnetic order. This leads to the question whether the Nernst anisotropy detects the same aspect of the nematicity as observed in the "one-directional" electrical resistivity and the thermoelectric power, and more importantly, whether the nematic fluctuations can be of magnetic origin? Up to now, several mechanisms were proposed as possibly relevant to the nematicity, including orbital [27], spin [28] or charge [29] order. Here we would like to point at reports indicating that the nematic and magnetic orders are distinct. For instance, studies of differential elastoresistance point at a nematic quantum critical point located inside the superconducting dome [4]. Results of polarization-resolved Raman spectroscopy studies link this quantum critical point to a charge quadrupole order that was shown to compete with the collinear antiferromagnetic order [30]. Another example is an emergence of two types of nematicity that was suggested to take place in the copper-based superconductor $YBa_2Cu_3O_y$ [31]

For the future, it will be interesting to perform similar anisotropy studies on cobalt doped $CaFe_2As_2$ compound, in which the magnetic/structural transition seems to be of the first order [32] and where magnetic fluctuations above $T_{tr}$ are absent or at least much weaker [22].

**Summary**

We studied the in-plane anisotropy of magneto-thermoelectric phenomena in a series of the $Ba(Fe_{1-x}Co_x)_2As_2$ single crystals. The uniaxial pressure was applied to the samples in order to detwin them in the orthorhombic phase and to allow the detection of macroscopic consequences of the emerging nematicity in the tetragonal phase. For all compositions studied the rotational symmetry of the electronic system is broken much above the temperature of the structural/magnetic transition. Moreover, the symmetry is broken even if such a transition is absent, as in the optimally doped sample Co6. We observe a strong thermoelectric anisotropy in the underdoped samples, whereas it is rather small and reversed in Co6. This change of the sign of anisotropy is attributed to the multiband structure of $Ba(Fe_{1-x}Co_x)_2As_2$, where the total transport coefficients are composed of contributions from different conduction bands tuned by the cobalt doping. An important, but to some extent unexpected result is the observation of an exponential temperature dependence of the Nernst anisotropy in the underdoped samples. Furthermore, $\Delta\nu$ turned out to be unaffected by the spin density wave transition, which suggests that nematicity in $Ba(Fe_{1-x}Co_x)_2As_2$ is distinct from the magnetic order.

**Acknowledgments**

This work was supported financially by the National Science Centre (Poland) under the research Grant No. 2014/15/B/ST3/00357.

**Figures**

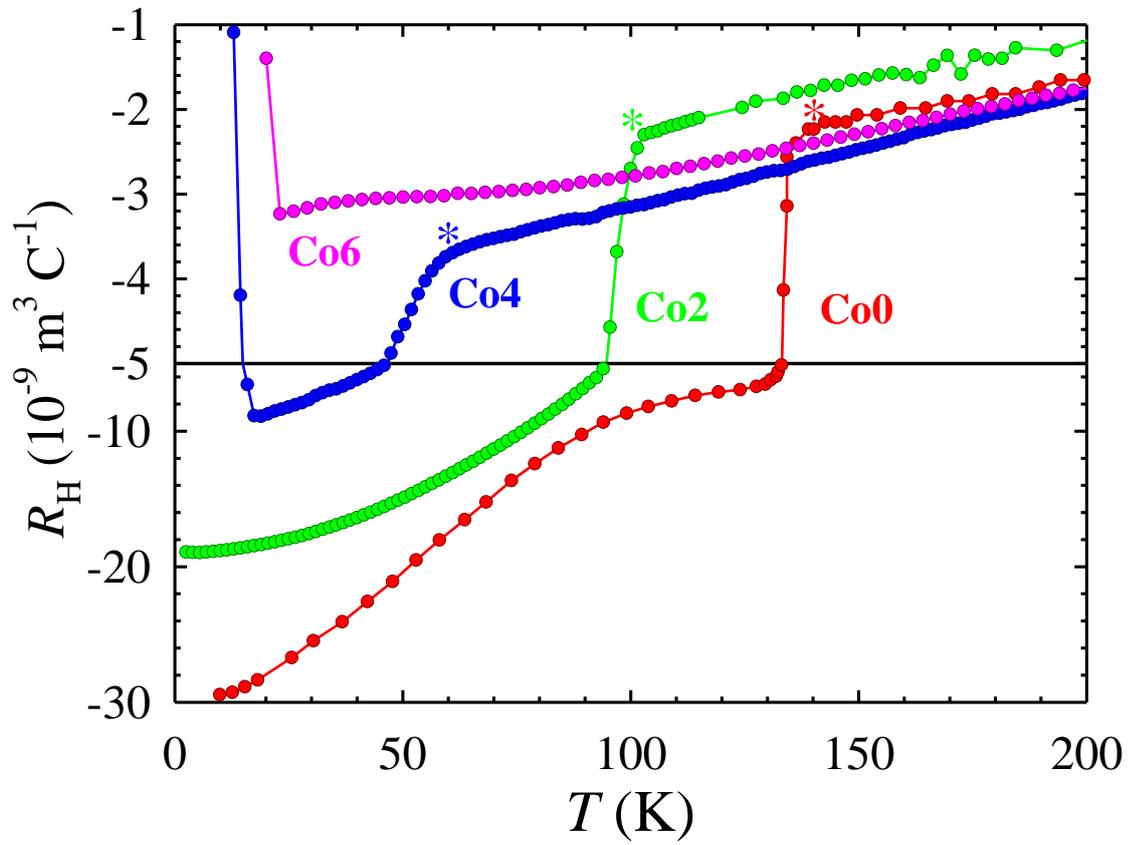

**Figure 1.**
(Color online) The temperature dependences of the Hall coefficient for the $Ba(Fe_{1-x}Co_x)_2As_2$ series (data for Co0 taken from Ref. [17]). Asterisks denote approximate temperatures of the magnetic/structural transitions. The top and bottom panels use different vertical scales.

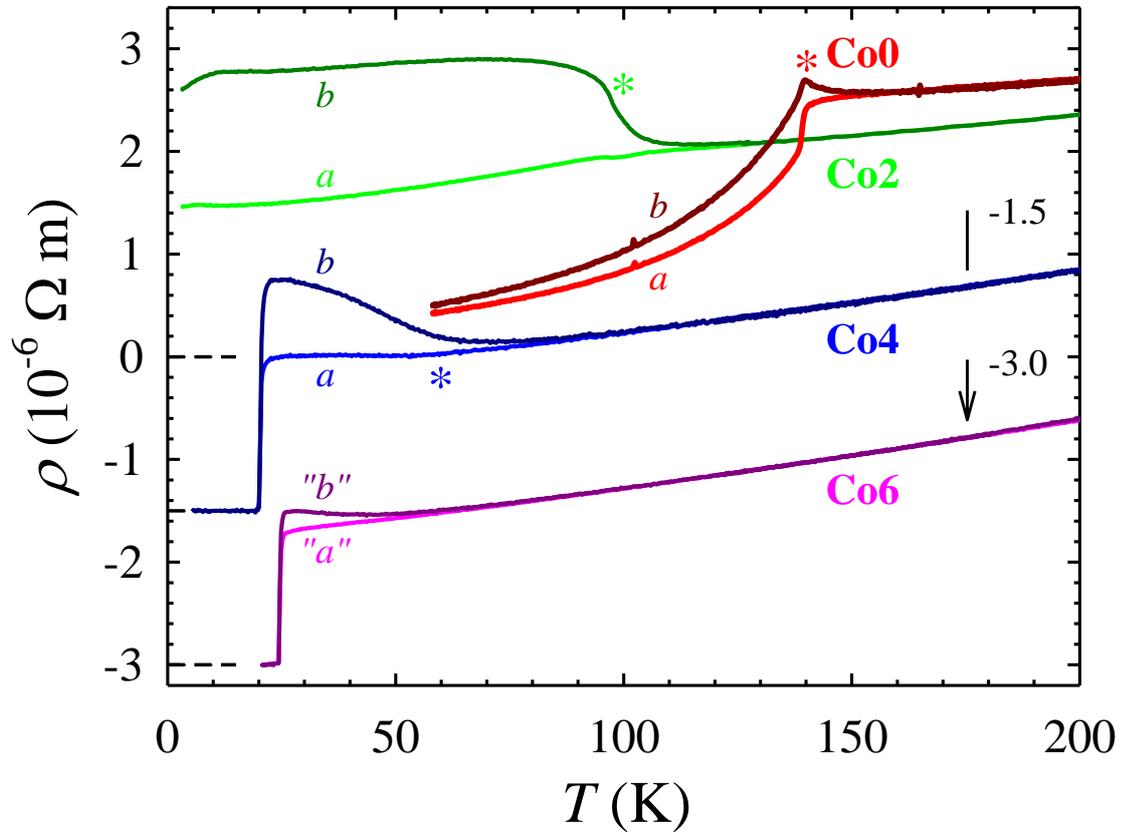

**Figure 2.**
(Color online) The temperature dependences of the resistivity for the Ba(Fe$_{1-x}$Co$_x$)$_2$As$_2$ series across (*a*) and along (*b*) the applied strain. Co2, Co4 and Co6 plots are shifted vertically for the sake of clarity. Asterisks denote approximate temperatures of the magnetic/structural transitions as taken from Fig. 1.

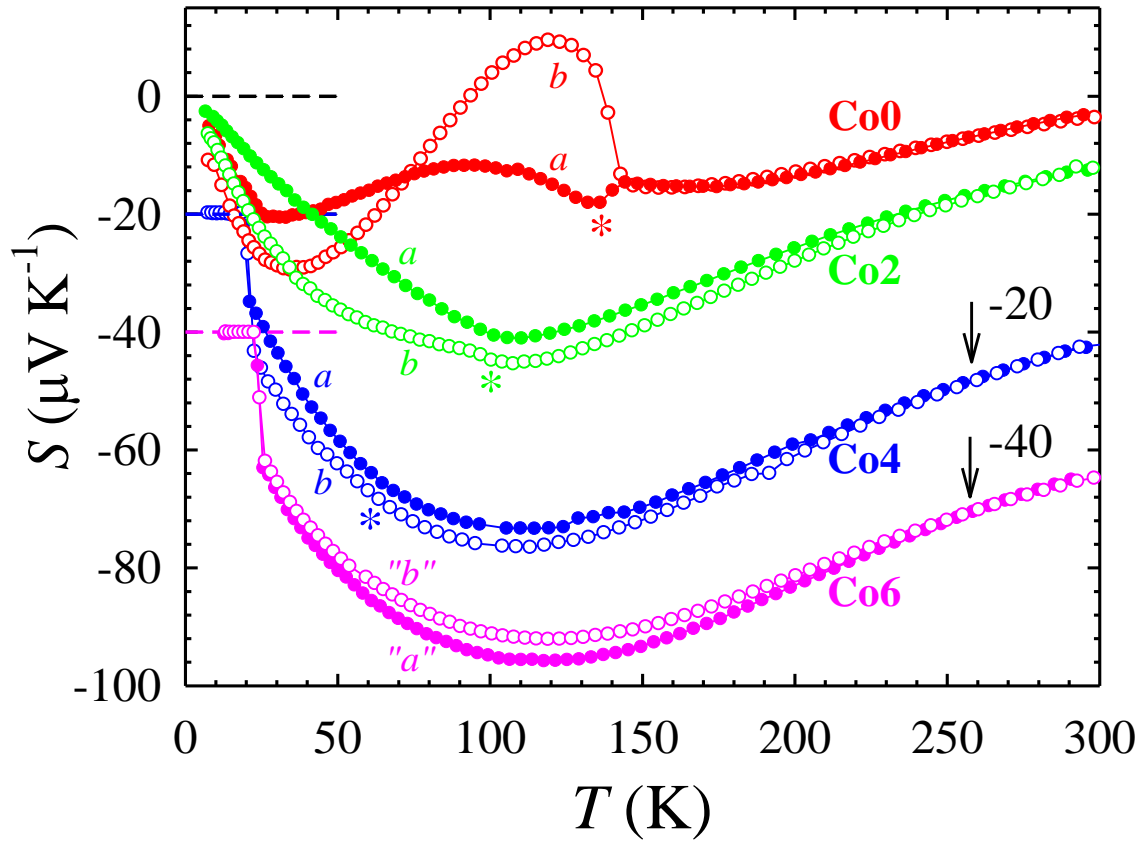

**Figure 3.**
(Color online) The temperature dependences of the thermoelectric power for the Ba(Fe$_{1-x}$Co$_x$)$_2$As$_2$ series across (solid points) and along (open points) the applied strain. Co2, Co4 and Co6 plots are shifted vertically for the sake of clarity. Asterisks denote approximate temperatures of the magnetic/structural transitions.

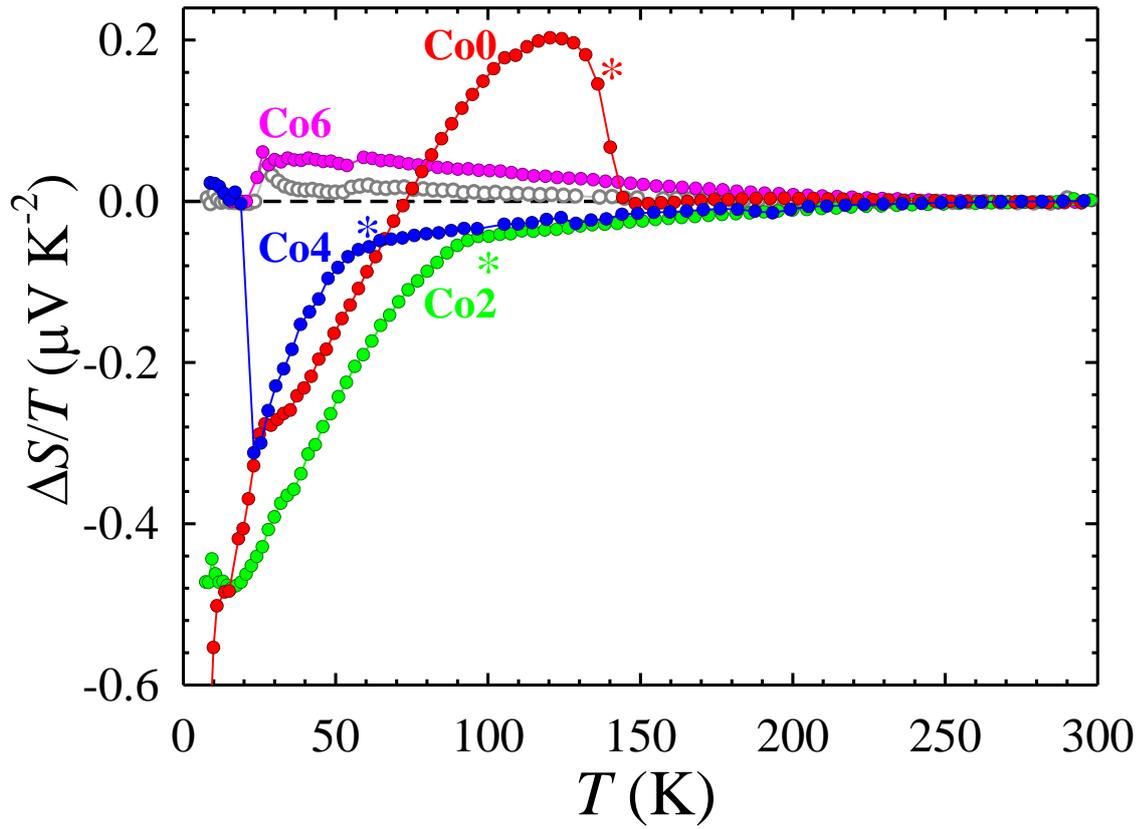

**Figure 4.**
(Color online) The temperature dependences of the Seebeck anisotropy $\Delta S/T = (S_b - S_a)T$ for the Ba(Fe$_{1-x}$Co$_x$)$_2$As$_2$ series. Asterisks denote approximate temperatures of the magnetic/structural transitions. Open points show the difference between the $S_b$ values measured for two different Ba(Fe$_{0.94}$Co$_{0.06}$)$_2$As$_2$ samples ($S_b^{\text{Co6\#1}} - S_b^{\text{Co6\#2}})/T$ providing a test of the reproducibility.

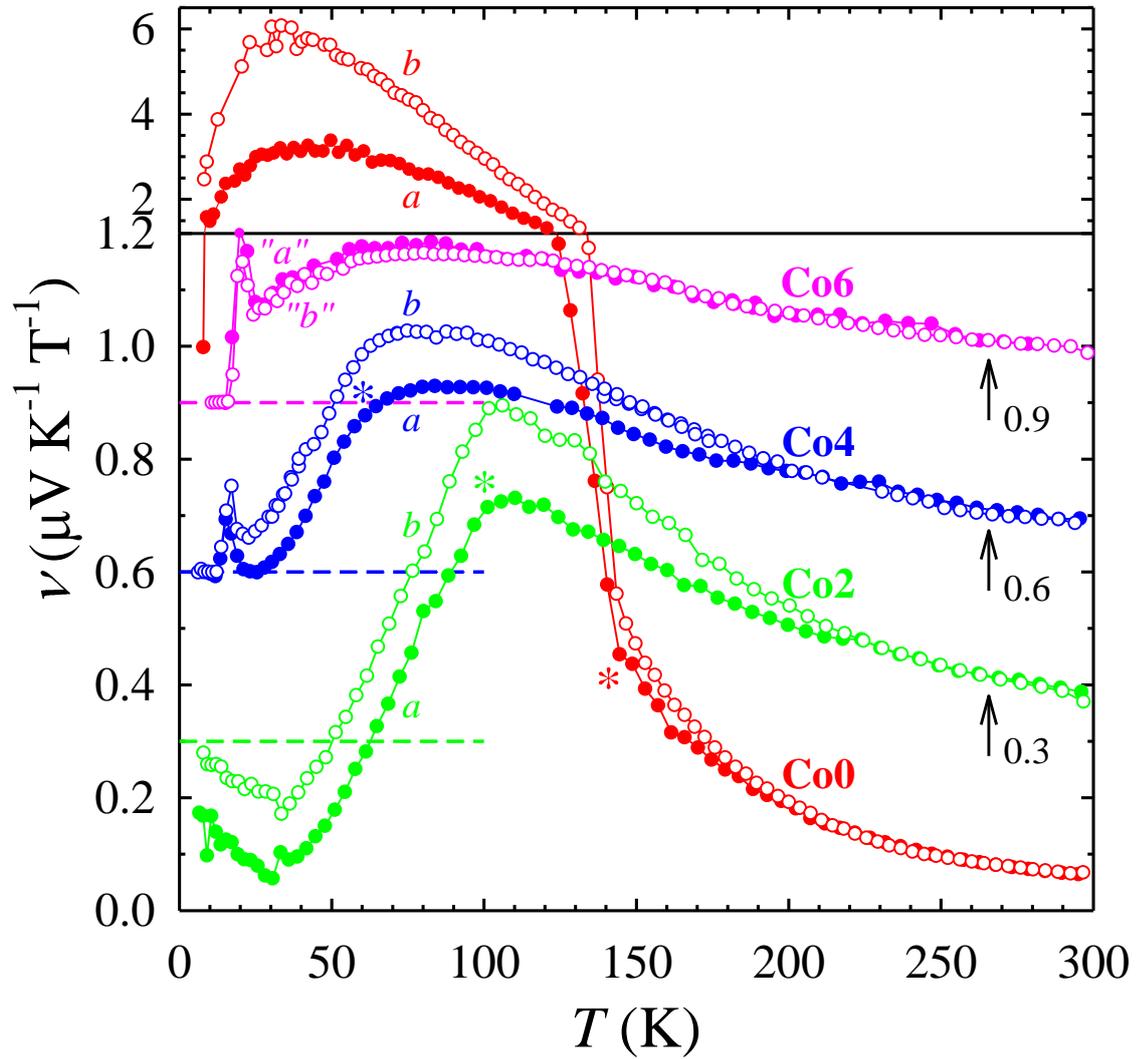

**Figure 5.**
(Color online) The temperature dependences of the Nernst coefficient in the Ba(Fe$_{1-x}$Co$_x$)$_2$As$_2$ series across (solid points) and along (open points) the applied strain. Co2, Co4 and Co6 plots are shifted vertically for the sake of clarity. Asterisks denote approximate temperatures of the magnetic/structural transitions. The top and bottom panels use different vertical scales.

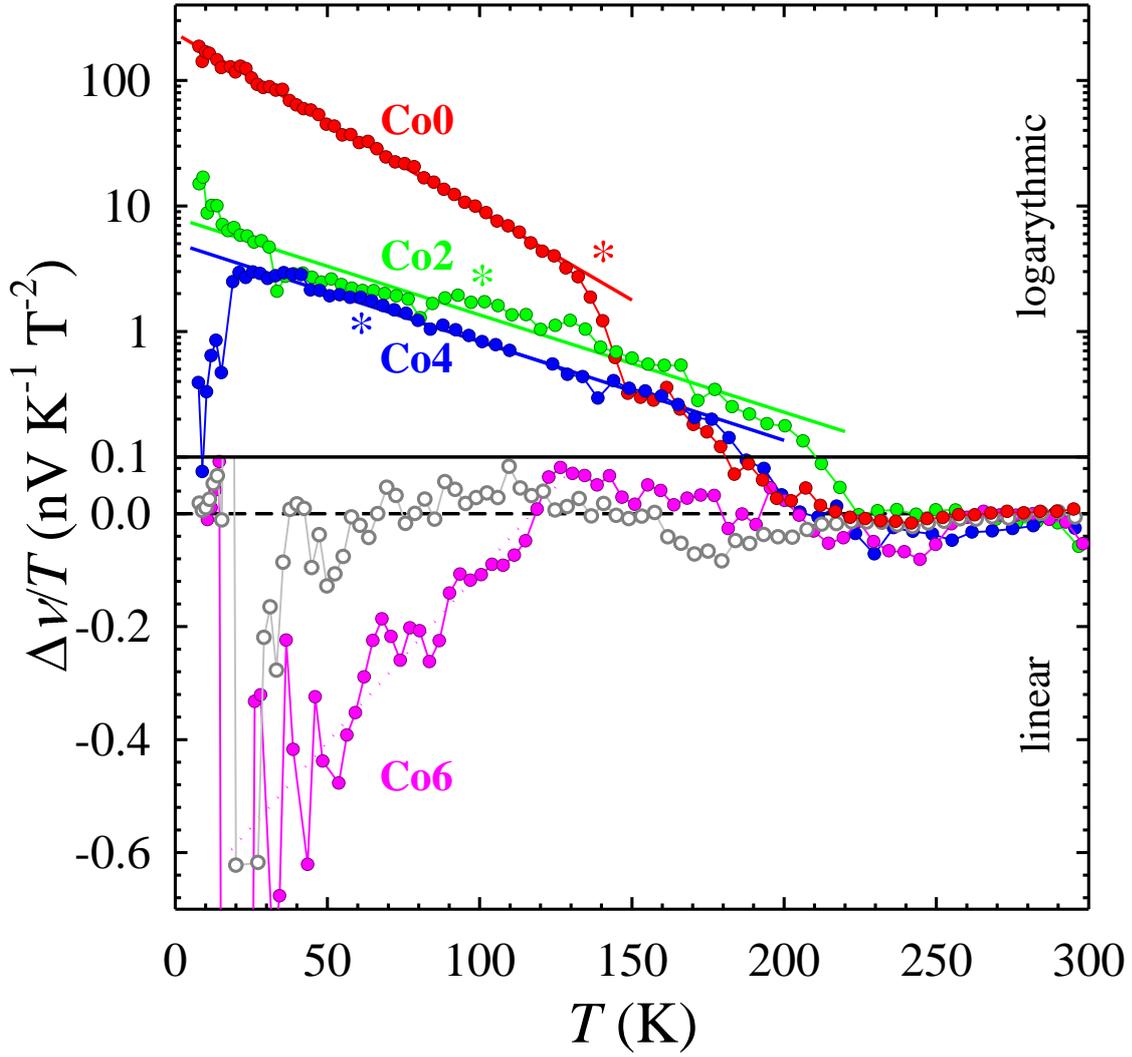

**Figure 6.**
(Color online) The temperature dependences of the Nernst anisotropy $\Delta v/T = (v_b\text{-}v_a)/T$ for the Ba(Fe$_{1-x}$Co$_x$)$_2$As$_2$ series. The solid lines represent $\Delta v(T)/T \sim e^{-cT}$ fits. Asterisks denote approximate temperatures of the magnetic/structural transitions. The top and bottom panels use different vertical scales. Open points show the difference between the $v_b$ values measured for two different Ba(Fe$_{0.94}$Co$_{0.06}$)$_2$As$_2$ samples ($v_b^{\text{Co6\#1}} - v_b^{\text{Co6\#2}})/T$ providing a test of the reproducibility.